# A simple EEG-based tool to guide therapeutic hypothermia decisions in neonatal hypoxic-ischemic encephalopathy

Marc Fiammante [a,b,*], Anne-Isabelle Vermersch [c], Marie Vidailhet [a,d], Mario Chavez [e]

[a] Paris Brain Institute Inserm U1127, CNRS UMR7225, Sorbonne Université UM75, Pitié-Salpêtrière Hospital, Inria Paris, Paris, France
[b] Retired IBM Fellow, France
[c] Physiology & Paediatric Functional Explorations Unit, Armand Trousseau Hospital, Paris, France
[d] Institut de Neurologie, Pitié-Salpêtrière Hospital, Paris, France
[e] CNRS UMR7225, Pitié-Salpêtrière Hospital, AP-HP, Paris, France

ABSTRACT

*Objective:* Accurate identification of hypoxic-ischemic brain injury in the early neonatal period is essential for initiating therapeutic hypothermia (TH) within 6 hours of birth to optimize neurodevelopmental outcomes. We aimed to develop a simple decision-making tool for identifying term neonates with hypoxic-ischemic encephalopathy (HIE) based on features of conventional electroencephalograms (EEG) recorded within 6 hours of birth.
*Methods:* EEG recordings from 100 full-term neonates with HIE were graded by pediatric neurologists for severity. Amplitude in slow frequency bands was analyzed, focusing on delta (0.5–4 Hz) spectral power. Temporal fluctuations of delta power characterized each HIE grade, with joint level and duration probability densities estimated for delta oscillation power. This study is registered on clinicaltrials.gouv (NCT05114070).
*Results:* These 2D EEG representations effectively distinguish mild HIE cases from those requiring hypothermia, achieving 98% accuracy, 99% sensitivity, 99% positive predictive value, 94% negative predictive value, an F1 score of 99%, and a false alarm rate of only 6%. This system accurately discriminates mild from moderate or severe HIE, with only one mild case mistakenly identified as requiring hypothermia and one moderate case erroneously flagged for treatment.
*Conclusions:* Quantized probability densities of delta spectral features from early EEG (within 6 hours of birth) revealed significant differences between mild and moderate/severe HIE, enabling accurate discrimination of candidates for TH.
*Significance:* Simple, interpretable biomarkers from early EEG can provide an efficient visual clinical decision support tool to identify full-term neonates with HIE eligible for therapeutic hypothermia.

## 1. Introduction

Perinatal asphyxia, defined as severe disruption of uteroplacental gas exchange leading to metabolic acidosis, can cause hypoxic-ischemic encephalopathy (HIE), the most common severe neurological morbidity in newborns [1,2]. HIE, as described by Sarnat and Sarnat [3], is diagnosed using clinical (e.g., consciousness, reflexes), biological (e.g., pH, lactate), and EEG markers (e.g., background rhythms, critical discharges). Early therapeutic hypothermia (TH) is the standard neuroprotective treatment for neonatal HIE [4]. Administered within the first six hours of life, TH reduces the risk of death or disability in moderate to severe HIE [5,6]. However, TH involves 72 hours of whole-body cooling in a neonate intensive care unit (NICU), often requiring deep sedation and assisted ventilation. Additionally, it can cause side effects, including heart rhythm and coagulation disorders, with potentially severe consequences [7].

Identifying neonates eligible for therapeutic hypothermia (TH) remains a clinical challenge, as rapid and accurate diagnosis is required within six hours of birth. Evidence clearly shows that the early analysis of EEG background provides useful information, beyond the clinical Sarnat score, to assess the presence and severity of encephalopathy [8]. While continuous EEG is a non-invasive bedside tool widely used to assess neonatal brain function, it requires specialized skills for recording and interpretation. Pediatric neurologists generally evaluate EEG features such as amplitude, frequency, continuity, and sleep-wake cycling [9]. Specific patterns, including low-voltage background activity, burst-suppression, prolonged inter-burst intervals, or theta-delta inversion, indicate brain injury. However, in many NICUs, TH decisions are often made without EEG criteria, limiting precise functional evaluation of cerebral electrical activity and hindering HIE grading and timely TH initiation [10]. For example, only 7 % of NICUs in France report 24/7 access to continuous EEG [11].

* Corresponding author.
*E-mail address:* marc.fiammante@icm-institute.org (M. Fiammante).

Unlike continuous EEG (cEEG), which provides detailed cortical dynamics, amplitude-integrated EEG (aEEG) offers a simplified trace derived from standard EEG and is widely used in neonatal care due to its accessibility and feasibility for long-term monitoring [12–14]. Although useful in evaluating neonatal encephalopathy, aEEG has notable technical and interpretive limitations, particularly in detecting mild to moderate hypoxic–ischemic encephalopathy [15,16]. These constraints underscore the importance of automated methods for analyzing cEEG, which can more accurately characterize background patterns and track encephalopathy progression.

Several algorithms have been proposed to identify full-term neonates with HIE eligible for therapeutic hypothermia (TH) based on continuous EEG analysis. Reliable EEG features have been explored across time, frequency, and information theory domains to quantify HIE severity [17–19]. Interburst intervals have been used to track encephalopathy progression [20–22], while nonstationary statistics, such as time-frequency decompositions [23,24] and long-range fluctuation analysis [25], have quantified EEG continuity as a marker of severity. Spatiotemporal dynamics of raw or parameterized EEG signals have also been effective for distinguishing HIE grades [24,26,27]. These features are commonly incorporated into machine learning approaches [28] or deep learning models [23,29] to develop automated grading systems for hypoxic-ischemic encephalopathy (HIE). Notably, the continuous assessment of cortical dynamics using a deep learning–based model has demonstrated promising results in predicting clinical outcomes [29,30].

Delta (0.5–4 Hz) and theta (4–8 Hz) oscillations dominate neonatal cortical activity compared to higher frequencies [31]. While the theta/delta power ratio serves as a marker of normal neonatal EEG, delta power is a robust biomarker for tracking encephalopathy severity [32,33]. Delta power can differentiate neonates with HIE who have significant brain injury from those with favourable outcomes [34].

This study proposes a straightforward quasi-real-time system utilizing conventional EEG analysis during the first six hours of life to detect brain injury and guide therapeutic hypothermia decisions in full-term neonates born with asphyxia. Our method, which leverages the combined information of delta power levels and their durations, effectively captures general power fluctuations in slow EEG oscillations, enabling accurate identification of neonates requiring hypothermia. By integrating temporal duration features with spectral power levels, this approach provides a statistically robust, interpretable, and practical decision-support tool for neonatologists in the NICU.

## 2. Methods

### 2.1. Dataset

This retrospective study included 104 neonates from a larger cohort of over 600 term neonates monitored between 2005 and 2020 at the NICU of Armand Trousseau Hospital (Paris, France). Fig. 1 illustrates our study population cohort. Inclusion criteria followed those of the French Society of Neonatology [35] and were: gestational age > 36 weeks, birth weight > 1800 g, postnatal age < 6 hours (H6), and HIE (mild, moderate, or severe) with at least one of the following features: (a) birth asphyxia with clinical signs (e.g., fetal distress, umbilical cord prolapse), (b) Apgar score ≤ 5 at 5 min, or (c) pH < 7 and/or lactates > 11 mmol/l in cord blood or within the first hour of life. Only patients with an EEG recording obtained during the first 6 hours of life were included in the cohort.

Among the patients analysed, 41 (including mild, moderate, and some severe cases) were not under hypothermia treatment at the time of EEG recording. Unfortunately, the timing of sedation administration was not systematically documented in our retrospective dataset and could not be reliably retrieved for all patients. No patients had received anti-epileptic medication. Details of patient exposure to the various sedation regimens are presented in Table 1. We notice however, that none of the patients were treated with ketamine or dexmedetomidine (Dexdor).

Parents of eligible patients were informed via a written document regarding anonymized data usage, and non-opposition consent was obtained. The study was registered with the French National Commission on Informatics and Liberty (CER-2021–023) and approved by the Sorbonne University Ethics Committee. It is also registered on clinicaltrials.gov (NCT05114070).

### 2.2. EEG recordings

EEG signals were recorded using Deltamed (2005–2016) and Nihon-Khoden (2016–2020) systems. Three patients' EEG recordings were excluded due to technical format issues. Continuous video EEG was obtained within the first six hours of birth after perinatal asphyxia to monitor encephalopathy progression. EEG data were recorded with Deltamed ($N = 27$) and Nihon-Khoden ($N = 77$) systems, sampled at 256 Hz and 500 Hz, respectively. High- and low-pass filters at 0.19 Hz and 70 Hz were applied during recording, and electrode impedance was kept below 10 kΩ. Physiological signals (ECG, respiration, SpO2) were recorded for some patients but were not used in this study. Although scalp EEG electrodes were placed in accordance with the international 10–20 system—using a reference electrode located at the midline between Fz and Cz—only the signals from electrodes Fp1, Fp2, T3, and T4 demonstrated consistency across all participants. Consequently, only these channels were included in the final analysis. EEG data from 4 patients were discarded due to file corruption ($n = 3$) or widespread artifact contamination ($n = 1$).

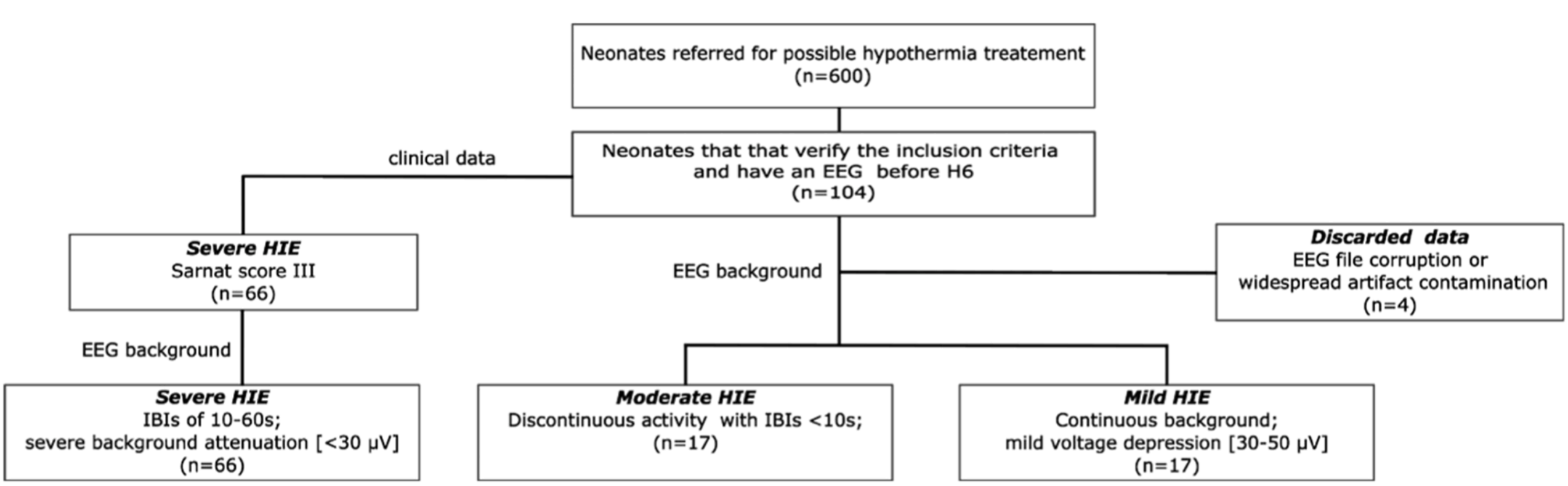

**Fig. 1.** Study population cohort diagram.

**Table 1**
Demographic characteristics of study group (median values [min-max]). Exposure to morphine (MOR), midazolam (MDZ), fentanyl (FEN), sufentanil (SUF), and phenobarbital (PB).

| Demographic data | Total | Severity groups | | |
|---|---|---|---|---|
| | | Mild HIE | Moderate HIE | Severe |
| Total (N) | 100 | 17 | 17 | 66 |
| Gender (M/F) | 60/40 | 9/8 | 9/8 | 42/24 |
| Gestational age at birth (weeks+days) | 39+3 [34+6 – 42+4] | 40+4 [36+4 – 41+6] | 39+3 [37+1 – 41+6] | 39+3 [34+6 – 42+4] |
| Apgar 1 min | 2 [0 – 8] | 2.5 [1 – 7] | 2.5 [0 – 8] | 2 [0 – 8] |
| Apgar 5 min | 6 [1 – 10] | 7 [3 – 10] | 5.5 [2 – 10] | 6 [1 – 10] |
| Umbilical cord gas PH | 7.04 [6.67 – 7.4] | 7.02 [6.8 – 7.32] | 7.06 [6.8 – 7.29] | 7.08 [6.67 – 7.4] |
| Lactates mmole/l | 10.8 [1.7 – 21] | 10.7 [3.4 – 16.5] | 11.55 [3.1 – 21] | 10.3 [1.7 – 20] |
| Sedation | | No sedation ($n = 15$), MDZ + MOR ($n = 2$) | No sedation ($n = 10$), MDZ + MOR ($n = 2$), MOR ($n = 5$) | No sedation ($n = 16$), MDZ+PB+FEN ($n = 1$), MDZ + FEN ($n = 13$), MDZ + MOR ($n = 19$), SUF + PB ($n = 1$), MOR + PB ($n = 1$), MDZ + SUF ($n = 1$), MOR ($n = 12$), FEN ($n = 2$) |
| Underwent TH | | 6 | 11 | 66 |
| EEGs performed during hypothermia | | 1 | 10 | 48 |
| Age at EEG recording (h) | 5 [2 6] | | | |
| Analysed EEG recording time (minutes) | 30.8 [5.8 – 56] | | | |

EEG signals were segmented into 1 s non-overlapping epochs. Segments containing significant artifacts, such as large motion or poor electrode contact, were visually identified and excluded from the analysis. Additionally, segments with alpha (8–12 Hz) spectral power exceeding $10^5$ $\mu V^2$/Hz were also considered as artefacts and removed. The proportion of excluded segments did not exceed 5 % of the total data during this artifact rejection process.

The patient groups were retrospectively defined according to clinical, biological, and neurophysiological criteria. The severe group was identified based on a combination of clinical manifestations and relevant biological parameters and confirmed by the visual analysis of EEG. In contrast, the moderate and mild groups were classified primarily according to the characteristics of the EEG background activity. Clinical evaluation using the Sarnat classification [3] remains an important tool, particularly when EEG is not available. A normal EEG provides an objective criterion for safely withholding therapeutic hypothermia. While our study was not designed to establish a detailed electro-clinical correlation, we consistently observed the following correspondences within our cohort: i) Sarnat grade I: normal EEG, no hypothermia, corresponding to mild HIE ; ii) Sarnat grade II: abnormal EEG, requiring hypothermia, corresponding to moderate HIE and iii) Sarnat grade III: markedly abnormal EEG, requiring hypothermia, corresponding to severe HIE.

The study included 100 newborns, categorized into three HIE grades: 17 with mild or minor encephalopathy (continuous background with mild voltage depression [30–50 μV]), 17 with moderate encephalopathy (discontinuous activity with IBIs <10 s), and 66 with severe encephalopathy (IBIs of 10–60 s and severe background attenuation [<30 μV]). The demographics of the study is summarized in Table 1.

### 2.3. EEG processing

Although originally developed for adult intensive care, aEEG is now widely used for continuous brain activity monitoring in NICUs. The signal is derived from EEG through asymmetric bandpass filtering (to reduce artefacts), rectification (to convert biphasic to monophasic waveforms), smoothing, and semi-logarithmic amplitude compression. Time compression then displays the maxima and minima of the transformed signal in a time window, enabling continuous monitoring and bedside interpretation over long recordings.

In this study, we developed a general representation of conventional EEG background dynamics by quantizing EEG power levels and measuring the durations at each power level. Focusing on slow EEG oscillations in the delta (0.5–4 Hz) band, we also assessed the theta/delta power ratio as a marker of normality in neonatal EEG. For each EEG segment, we quantified delta power and the theta/delta ratio, applying a logarithmic transformation ($Log_{10}$), and quantizing values to the first decimal. This approach preserves small changes while preventing overloading from large values. Pivot tables were created to capture changes in quantized power levels (or power ratios) across EEG segments, using the sequence of levels and durations. These tables were normalized to create a joint duration/level probability distribution. Gaussian smoothing was then applied to obtain 2D smooth probability densities.

The general scheme of our method is illustrated in Fig. 1. For all EEG recordings, bipolar channels Fp1-T3 and Fp2-T4 were used, and power spectral density was estimated for each 1-second segment using the Welch periodogram method.

### 2.4. HIE grade severity classification

We first evaluated the ability of level-duration distributions to distinguish between different HIE grades using a nearest-neighbor classifier. A leave-one-out cross-validation was applied: for each fold, one EEG recording was randomly selected for testing, while the others were used for training. The average probability distribution for each grade was estimated from the training set. The 2D level-duration representation of the test data was then compared with the reference distributions, and the nearest category (grade) was assigned based on the closest match. To compare two probability distributions $P_i$ and $P_j$ on sample space $\xi$, we computed the $L_2$ distance $d(P_i, P_j) = \sqrt{\sum_{x \in \xi} |P_i(x) - P_j(x)|^2}$ [36].

We evaluated our classifier's ability to distinguish the mild HIE group from those requiring hypothermia (severe and moderate HIE merged into a single class). Classification performance was assessed using precision (positive predictive value, the proportion of true positives among predicted positives), recall (sensitivity, the ability to identify all positives), negative predictive value and false positive rate (false alarm ratio). To mitigate the impact of class imbalance—particularly due to the predominance of severe HIE cases—we additionally reported balanced accuracy (the average of sensitivity and specificity) and the F1-score (the harmonic mean of recall and precision).

## 3. Results

### 3.1. Results on the French dataset

Fig. 2 displays the delta power fluctuation densities for mild, moderate, and severe HIE categories. These 2D representations, created for each patient's EEG signals, reveal distinguishable shape differences across HIE grades that are easily recognizable. Interestingly, the plots in Fig. 2 indicate that different HIE grades can be distinguished by power variations occurring over short time durations, patterns that standard spectral EEG features fail to capture.

As an initial assessment of the discriminative ability of our method, we computed the intra- and inter-group distances between the distributions of the three HIE categories (Table 2). In all cases, intra-group distances were consistently smaller than inter-group distances, confirming the robustness of the approach in separating HIE severity levels. The largest separation was observed between the mild and severe groups, while moderate cases occupied an intermediate position, as expected.

A cross-validation procedure proved our classification model consistency. Indeed, the classification of EEG segments correctly matched the grade in 81 % of cases: 53 % severe, 16 % mild, and 12 % moderate. Misclassifications included one mild graded as moderate, four moderate as severe, and thirteen severe as moderate. Interestingly, no mild cases were classified as severe or vice versa. In differentiating infants with mild HIE from those requiring therapeutic hypothermia (moderate or severe HIE), the classifier achieved consistently high performance metrics: recall (sensitivity) 99 % [95 % CI: 96–100], precision 99 % [95 % CI: 96–100], negative predictive value 94 % [95 % CI: 82–100], balanced accuracy 96 % [95 % CI: 90–100], false alarm ratio 6 % [95 % CI: 0–16], and F1-score 99 % [95 % CI: 97–100]. The narrow bootstrap-derived confidence intervals underscore the robustness and reliability of the model's discriminative ability. Clinically, misclassification was minimal, with only one mild case incorrectly identified as requiring hypothermia and one moderate case erroneously flagged for treatment.

**Table 2**
Minimum, mean, and maximum distances within and between HIE categories.

| Reference class | Category | Min | Mean | Max |
|---|---|---|---|---|
| Mild | Mild | 0.008 | 0.023 | 0.048 |
| | Moderate | 0.007 | 0.044 | 0.075 |
| | Severe | 0.062 | 0.082 | 0.102 |
| Moderate | Mild | 0.020 | 0.045 | 0.081 |
| | Moderate | 0.011 | 0.028 | 0.068 |
| | Severe | 0.034 | 0.066 | 0.091 |
| Severe | Mild | 0.024 | 0.088 | 0.157 |
| | Moderate | 0.013 | 0.071 | 0.157 |
| | Severe | 0.009 | 0.046 | 0.128 |

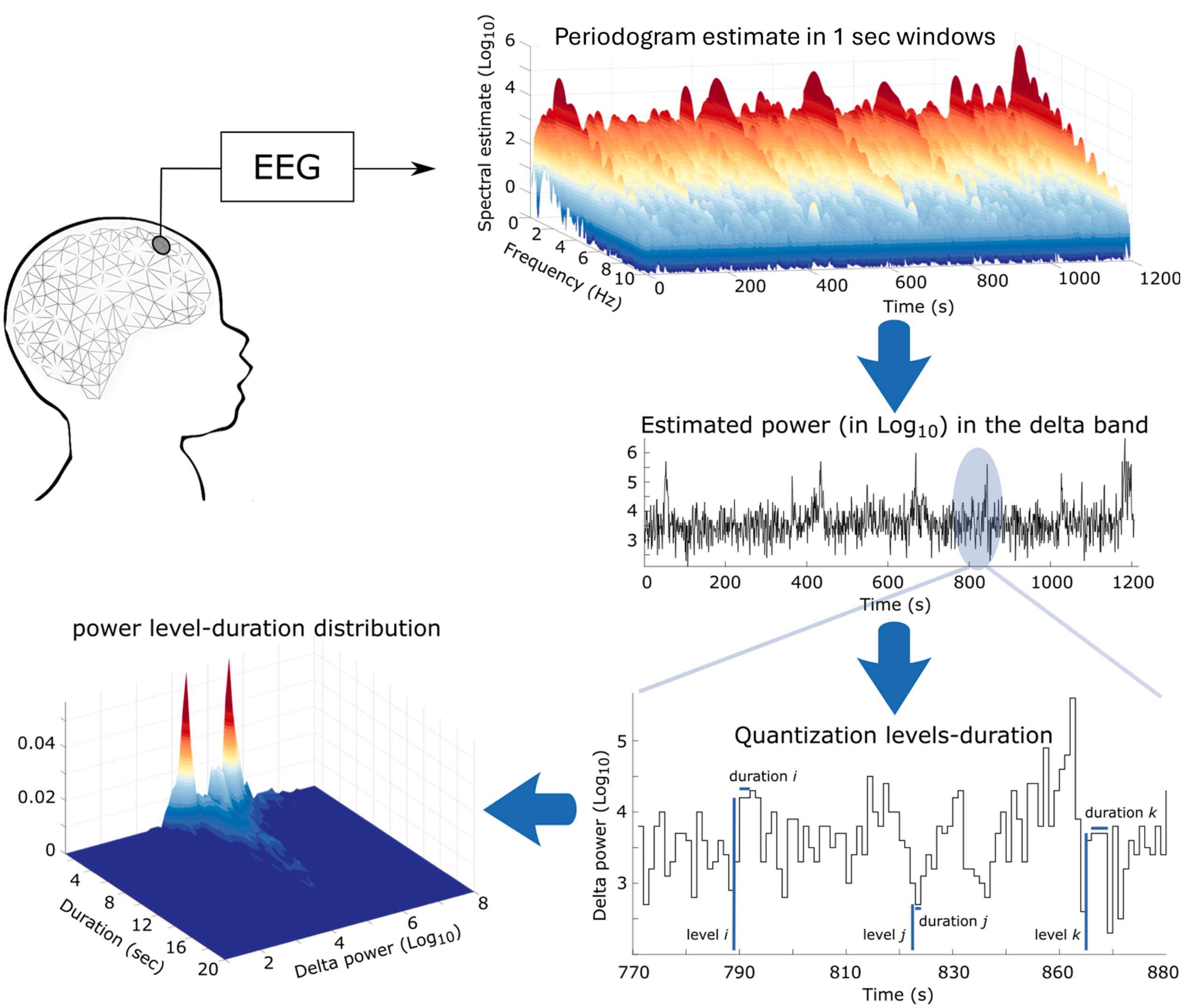

**Fig. 2.** Overview of the proposed feature extraction for each EEG channel, focusing on delta power. The plotted twenty-minute signal is shown for illustration purposes only and does not represent the segmentation applied to all EEG recordings.

To assess the robustness of our model when applied to shorter EEG segments, we reanalysed the data using resampled recordings of a fixed 10-minute duration. One patient with severe HIE was excluded due to a shorter available recording. The performance metrics are: recall (sensitivity) 98 % [95 % CI: 94–100], precision 98 % [95 % CI: 94–100], negative predictive value 88 % [95 % CI: 71–100], balanced accuracy 93 % [95 % CI: 85–99], false alarm ratio 11 % [95 % CI: 0–26], and F1-score 98 % [95 % CI: 95–99]. These results demonstrate that the use of a reduced and fixed-length EEG epochs yields consistent and robust classification performance, confirming the reliability of our approach across different EEG duration settings.

Other spectral features can also serve as markers of normality in neonatal EEG. When compared with the delta power, the level-duration distribution of the theta/delta spectral ratio showed moderate performance in discriminating between EEG segments of the mild HIE group and those requiring hypothermia. Performance metrics included sensitivity of 88 % [95 % CI: 83–93], precision 95 % [95 % CI: 89–99], and an F1-score of 91 % [95 % CI: 87–95]. Balanced accuracy was 82 % [95 % CI: 71–93], and the false alarm ratio remained moderate at 23 % [95 % CI: 6–42]. In contrast, the negative predictive value was lower at 57 % [95 % CI: 35–78], reflecting a reduced ability to reliably exclude cases. When using a reduced but fixed recording duration of 10 min, classification performance remained stable, with a sensitivity of 87 % (95 % CI: 81–92), precision of 95 % (95 % CI: 89–99), and an F1-score of 90 % (95 % CI: 86–94). The balanced accuracy was 81 % (95 % CI: 70–92), while the negative predictive value decreased to 54 % (95 % CI: 33–75), and the false alarm ratio increased slightly to 24 % (95 % CI: 6–43). These findings indicate high discriminative capacity, particularly in correctly identifying positive cases (including for shorter EEG recordings), but highlight the need for cautious interpretation of negative predictions in clinical practice.

### 3.1. Results from a public EEG-HIE dataset

To ensure clarity and reproducibility, we evaluated our method on a publicly available dataset from Cork University Maternity Hospital, Ireland [37]. This dataset includes over 100 multichannel EEG files recorded from 53 full-term newborns diagnosed with HIE. In the absence of pre-frontal electrodes, we analyzed signals from the frontal regions (F3 and F4). Each 1-hour EEG recording was graded for background abnormalities. Unlike our EEGs, recorded within 6 hours of birth, this dataset includes recordings up to 100 hours after birth. It is worthy noting that the Cork dataset uses a four-grade HIE classification: normal or mildly abnormal, moderately abnormal, severely abnormal, and inactive. Additionally, the version of the dataset used in this study (May 2022 release of [37]) does not include labels indicating whether neonates required therapeutic hypothermia.

Applied to this dataset, classification based on delta power resulted in a direct grade match for 77 % of EEGs. The remainder showed adjacent grade matching: 9.6 % of normal cases were graded as moderately abnormal, 1.9 % of moderately abnormal cases were misclassified as severely abnormal, and 0.95 % of severe cases were erroneously classified as inactive.

Although other statistical distances, such as Kullback-Leibler, Hellinger, or Bhattacharyya [36], could be used to compare probability distributions, our approach with the simple $L_2$ norm generally outperforms these alternatives. For example, using the Bhattacharyya distance only increased accuracy by less than 3 %.

## 4. Discussion

From a large cohort of full-term neonates with mild, moderate, and severe HIE, annotated clinical characteristics, and conventional EEGs (recorded before 6 hours), we demonstrated that spectral fluctuations in EEG background (summarized by delta power levels and their duration) can effectively discriminate between HIE categories. We validated our algorithms on an independent, publicly available EEG-HIE cohort [37], including normal term newborns and those with mild, moderate, and severe HIE. Additionally, we provided visual representations of HIE severity, easily interpretable by clinicians. Our approach offers a simple and efficient clinical decision support tool to identify full-term neonates with HIE eligible for therapeutic hypothermia.

Our results align with previous studies using quantitative EEG in full-term neonates and support the use of simple, automated EEG analysis as an accessible, generalizable method for identifying interpretable biomarkers of brain injury in neonates with HIE. Notably, the 2D level-duration probability distributions for each EEG effectively capture characteristics of different HIE grades, easily recognizable by clinicians. For the mild grade, the distribution shape clearly distinguishes between awake and sleep states. The two peaks reflect high-voltage and low-voltage waveforms during quiet sleep, while the variability of awake EEG is captured in the central values. The rightward shift in the distribution for moderate HIE indicates reduced EEG amplitude. Finally, the 2D density for severe HIE distinctly shows discontinuous, low-amplitude EEG patterns compared to the mild group.

A previous study using features from quadratic time-frequency decompositions achieved a classification accuracy of 78 %, which could be improved by incorporating sleep state information [24]. Similar performance (AUC $\approx$ 79 %) was reported by Kota et al [33] using delta power. Improved classification accuracy (83 %) was achieved by combining a large feature set with probabilistic models (Gaussian Mixture Models) and supervised machine learning (Support Vector Machines, SVM), followed by post-processing [28]. Similarly, Montazeri et al [29] reported classification accuracy over 90 % using a large feature set with different classifiers (SVMs and neural networks). More complex grading systems based on deep learning methods have achieved very good accuracy (>85 %) with various EEG features [23,38]. Our results show that even though the classifier based on the closest Euclidean distance to the reference distributions (obtained from the learning set) is one of the simplest and most common, its performance competes with more complex classifiers. It is important to acknowledge that accuracy metrics may be biased toward the majority class, particularly patients with unfavorable outcomes. To address this limitation, our study incorporates the F1-score and balanced accuracy, which are better suited to account for class imbalances and provide a more reliable assessment of model performance.

Our findings highlight the novelty and effectiveness of using a joint 2D level-duration distributions to capture spectral fluctuations of delta power from EEG background, enabling reliable discrimination between mild and more severe HIE grades. This approach achieved an accuracy of 98 %, with only 1 % of mild cases misclassified as requiring hypothermia. When tested on a public dataset with four HIE grades, the system maintained a robust overall classification accuracy of 77 %, with minimal misclassification: only 0.19 % of moderately abnormal EEGs were misclassified as severely abnormal, and 2.8 % of severely abnormal or inactive EEGs were misidentified as normal or mild. This demonstrates our approach's consistency, efficiency, and good performance across multiple datasets.

It is important to note that previous studies mentioned above were conducted on datasets with four HIE groups. However, our findings are consistent with previous classification results on datasets with three grades (mild, moderate, and severe). In Matic et al [26], the three HIE severity groups were correctly identified with an accuracy of 89 % using a complex system that integrated spatial EEG information for supervised classification. A recent study [39] reported achieving a classification accuracy of 80 % using a combined feature set optimized with an SVM classifier. In contrast, our approach offers a novel and efficient solution, achieving a direct matching of 81 % in our dataset and 77 % in the public EEG database, demonstrating superior simplicity and speed compared to existing methods.

Discriminating adjacent HIE grades (e.g., mild-moderate, moderate-severe) has proven challenging [28,26]. However, in our dataset, only 5.8 % of mild cases were misclassified as moderate, and vice versa. When applied to a public dataset with four HIE grades, our method misclassified 9.5 % of moderately abnormal EEGs as severely abnormal. For distinguishing the mild-normal HIE group from those requiring hypothermia, our approach achieved high accuracy, with only two misclassifications (2 %) in our dataset. This is particularly important, as clinical trials indicate that therapeutic hypothermia (TH) is most beneficial for neonates with moderate HIE. While accurate detection of normal or mild EEGs is vital for confirming a good clinical outcome, precise identification of moderate HIE cases is crucial for selecting neonates eligible for TH. Our method demonstrated a good accuracy in detecting normal or mild encephalopathy, with 94 % matching, highlighting the potential utility of this approach in guiding early therapeutic decisions

Our two-dimensional representation provides a comprehensive view of EEG background dynamics by jointly quantifying the duration of spectral power at different levels within the delta band. As shown in Fig. 3, neither high-power nor low-power oscillations alone reliably distinguish all HIE grades. In contrast, our method integrates power level and duration to accurately identify mild HIE cases and differentiate them from moderate and severe grades, which require neuroprotective therapy. Although the level–duration distribution can also be derived from the theta/delta spectral ratio, its performance in identifying the clinically relevant HIE group was moderate, with an accuracy of 86 %.

Unlike complex feature sets or classification algorithms, our system, which uses a single spectral feature and a simple classifier, achieves comparable classification performance. The clinical interpretability of EEG features is crucial for developing informative biomarkers of neurological injury in neonates with HIE. We believe that our 2D representation of EEG offers neonatologists a traceable method for identifying full-term neonates with HIE who may benefit from therapeutic hypothermia, without the “black box” limitations of more complex deep learning models [23,29] or undisclosed algorithms used in automated grading systems [40,41].

Due to the variety of medications and the relatively small group sizes, it is challenging to draw definitive conclusions regarding their effects on background EEG activity in our study. A comprehensive statistical assessment of the potential confounding effects of all administered medications across all HIE severity grades was not possible to perform. Although sedative medication could influence EEG background activity, infants across moderate and mild HIE groups were exposed to similar regimens of sedation and/or analgesia. Therefore, medication is unlikely to explain the observed differences between these two cases.

While promising, our study has limitations. First, the HIE severity was retrospectively assessed in a dataset of limited sample size. Future studies with larger, prospectively graded patient cohorts are needed to validate our approach. Our cohort consisted predominantly of infants with severe HIE, with only a small proportion of mild cases. This distribution, inherent to the retrospective nature of the study, may limit the generalizability of our findings to more representative clinical populations and could lead to overestimation of some performance metrics. Second, given the heterogeneity of electrodes location among patients in the French dataset, we cannot generalize the model to other derivations. Third, EEG grade can evolve during long-term monitoring due to the dynamic nature of HIE. A clinical decision tool should incorporate this variability to create a more flexible automated grading system [23].

Although EEG provides an objective assessment of encephalopathy severity, the absence of complementary neurological examination data (e.g. Sarnat or Thompson score) may limit direct comparison with clinical grading systems used in most centers. Finally, our results were based on preselected data with EEG epochs free of major artifacts. Incorporating advanced artifact detection methods [42,43] could improve the robustness and informativeness of the decision support system. To expand the clinical utility of our findings, future work should evaluate our method on larger datasets without preselecting clean EEG segments.

## 5. Conclusion and future work

Accurately identifying neonates with HIE who require therapeutic hypothermia is crucial for optimizing outcomes. The proposed approach explicitly separates the amplitude and temporal characteristics of EEG oscillations, generating intuitive “fingerprints” of HIE severity with distinct and recognizable patterns (e.g., bimodality in mild HIE, rightward shift in moderate HIE, and low-amplitude discontinuity in severe HIE). These two-dimensional representations are intended as an adjunct to support interpretation rather than as a stand-alone clinical score. Nevertheless, future work could derive additional automated or simplified quantitative indices from these visualizations to further facilitate bedside application.

Given the variability in electrode placements across different NICUs, a single universal discriminative model is not feasible. Our proposed approach utilizes data from pre-frontal regions (or frontal electrodes in the Cork dataset). However, HIE discrimination can be enhanced by combining information from multiple EEG channels [24,26,38], or by developing more complex models that account for the spatio-temporal organization of EEGs [44,45]. Despite this, such systems have limited clinical applicability as neonatal EEG recordings in NICUs typically involve only a few channels, many of which may require rejection due to artifacts. More sophisticated and accurate decision support systems could be achieved by integrating clinical markers [46], data from functional near-infrared spectroscopy [47], or other physiological signals like heart rate variability [48] to further enhance clinical decision-making [49].

While larger patient cohorts are needed to fully assess the limits of our approach, our results indicate that the proposed method offers a robust, simple, interpretable and automated tool for assessing brain injury in neonates with HIE. The spectral power fluctuations in other frequency bands could also aid in detecting sleep-wake transitions or tracé alternant activity during quiet sleep in neonates. Our system could enable clinical and nursing staff to make objective, data-driven decisions about whether to initiate hypothermia treatment based on the presence or absence of brain injury. This easy-to-use decision support tool could reduce delays in starting treatment when necessary, preventing the administration of unnecessary treatments and ensuring timely hypothermia for eligible neonates. Ultimately, this approach has the potential to minimize neurological disabilities by providing critical treatment in a timely manner. Deployed in NICUs, it offers a promising solution to reduce the number of children and adults living with disabilities.

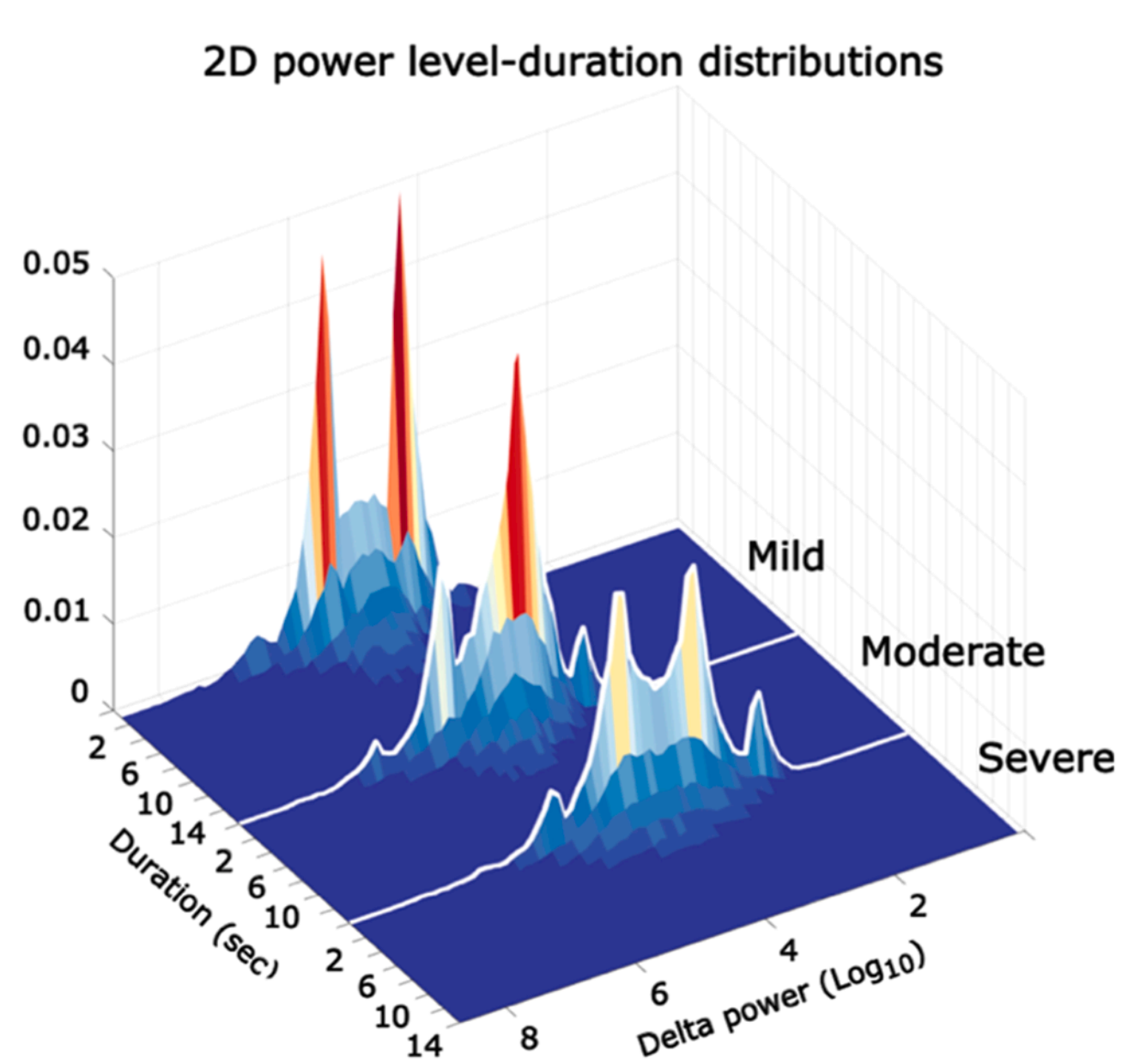

**Fig. 3.** Joint level-duration probability densities of delta power for the three HIE grades. The normalized densities represent the average densities obtained from all mild, moderate, and severe cases.

## Ethics statement

This research was conducted in accordance with the ethical standards of the relevant national and institutional committees. This study is registered on clinicaltrials.gouv (NCT05114070).

## CRediT authorship contribution statement

**Marc Fiammante:** Writing – original draft, Software, Investigation, Conceptualization. **Anne-Isabelle Vermersch:** Writing – review & editing, Resources, Project administration, Funding acquisition, Data curation, Conceptualization. **Marie Vidailhet:** Writing – review & editing, Conceptualization. **Mario Chavez:** Writing – review & editing, Visualization, Supervision, Resources.

## Declaration of competing interest

The authors declare the following financial interests/personal relationships which may be considered as potential competing interests: Marc Fiammante, Mario Chavez and Anne-Isabelle Vermersch have patent METHOD FOR PROCESSING AN ELECTROENCEPHALOGRAM SIGNAL pending to ASSISTANCE PUBLIQUE - HÔPITAUX DE PARIS European Parent Office Application number EP24305357.6. If there are other authors, they declare that they have no known competing financial interests or personal relationships that could have appeared to influence the work reported in this paper.

## Acknowledgments

Early stage of this work was supported by the Human Safety Net (Generali Fundation). Charitable Donation Ref N° 171437A20 (2019).